\documentclass[aps,twocolumn,showpacs,amsmath,amssymb,10pt]{revtex4-1}
\usepackage{graphicx}
\usepackage{subfigure}
\usepackage{ulem}
\usepackage{amsmath}
\usepackage{amssymb}
\usepackage{wrapfig}
\usepackage[utf8]{inputenc}  
\usepackage{color} 

\begin{document}

\title{Measurement of van-der-Waals interaction by atom trajectory imaging}
\author{N.~Thaicharoen*}
\author{A.~Schwarzkopf$^{\dag}$}
\author{G.~Raithel}
\affiliation{ Department of Physics, University of Michigan, Ann Arbor, MI 48109}
\date{\today}

\begin{abstract}
We study the repulsive van der Waals interaction of cold rubidium $70S_{1/2}$ Rydberg atoms by analysis of time-delayed pair correlation functions. After excitation, Rydberg atoms are allowed to accelerate under the influence of the van der Waals force. Their positions are then measured using a single-atom imaging technique. From the average pair correlation function of the atom positions we obtain the initial atom-pair separation and the terminal velocity, which yield the van der Waals interaction coefficient $C_{6}$. The measured $C_{6}$ value agrees well with calculations. The experimental method has been validated by simulations. The data hint at anisotropy in the overall expansion, caused by the shape of the excitation volume. Our measurement implies that the interacting entities are individual Rydberg atoms, not groups of atoms that coherently share a Rydberg excitation.
\end{abstract}

 %

\maketitle
	
The van der Waals interaction is important in the description and control of interactions in few- and many-body dynamics studies. This interaction has been critical in the observation of Rydberg excitation blockades and collective excitations~\citep{lukin_dipole_2001, tong_local_2004, liebisch_atom_2005, garttner_collective_2014}, Rydberg crystals~\citep{pohl_dynamical_2010, schaus_dynamical_2014}, and Rydberg aggregates~\citep{lesanovsky_out--equilibrium_2014, urvoy_strongly_2015}. Rydberg interactions have been used in quantum information processsing~\citep{isenhower_demonstration_2010, wilk_entanglement_2010, keating_adiabatic_2013, petrosyan_binding_2014}. The van der Waals interaction between two Rydberg atoms has been measured using spectroscopic methods~\citep{reinhard_double-resonance_2008,beguin_direct_2013}. Several measurements have been performed near surfaces to observe radiative Rydberg-level shifts caused by image charge interaction near metal surfaces~\citep{sandoghdar_direct_1992, nordlander_interaction_1996}.  The van der Waals interaction between excited cesium atoms and a dielectric surface has been measured using selective reflection spectroscopy~\citep{fichet_exploring_2007}.
		
\begin{figure}[b]
	\centering
			\includegraphics[width=1\linewidth]{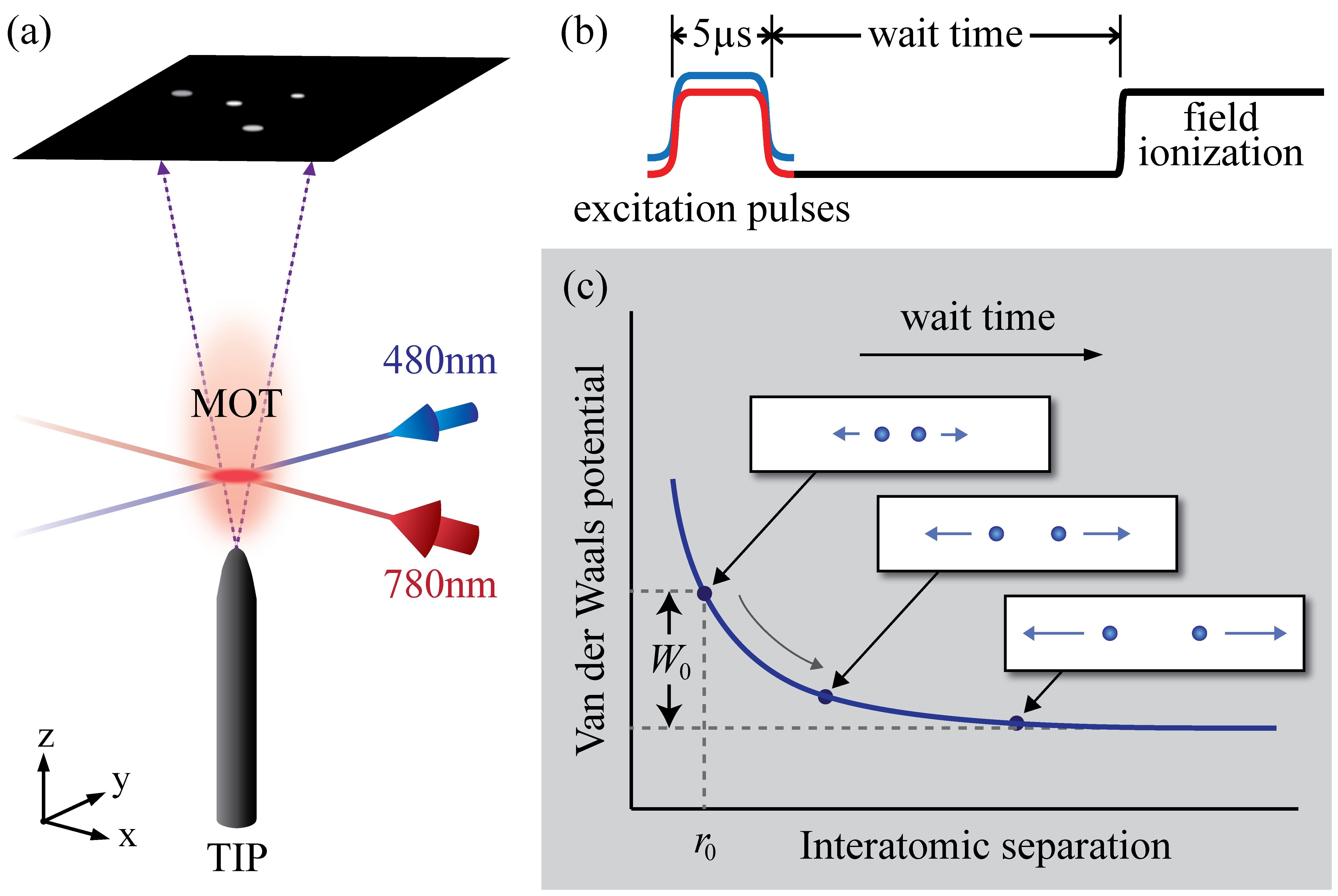}
	\caption{(Color online) Geometry (a) and timing sequence (b) of the experiment. The wait time is the time difference between the end of the excitation and the onset of the field ionization pulse. (c) Van der Waals potential versus separation between two Rydberg atoms.  The separation increases due to the repulsive van der Waals force. We image the resulting trajectories by varying the wait time. In the insets in (c), circles represent two Rydberg atoms evolving under the influence of the van der Waals force. The (final) kinetic energy release, indicated by the blue velocity arrows, equals the van der Waals potential $W_{0}$ at the initial separation $r_{0}$.}
	\label{fig:exp_setup}
\end{figure}

Here, we develop a method to study the van der Waals interaction between Rydberg atoms using direct spatial imaging of their trajectories~\citep{schwarzkopf_imaging_2011, schaus_observation_2012, mcquillen_imaging_2013, schwarzkopf_spatial_2013}.  Pairs of $70S_{1/2}$ rubidium Rydberg atoms are prepared with a well-defined initial separation by detuning an excitation laser and utilizing the $r^{-6}$ dependence of the van der Waals interaction~\citep{robicheaux_ionization_2005, schwarzkopf_spatial_2013}. After preparation, the atoms are subject to van der Waals forces (which are repulsive in this case). The effect of the forces is observed by tracking the interatomic distance between the Rydberg atoms, after they have been allowed to move for selected wait times (see Fig.~\ref{fig:exp_setup}). The atom trajectories and thereby the van der Waals interaction coefficient $C_{6}$ are extracted from the pair correlation functions of the Rydberg atom positions.

The experimental setup is shown in Fig.~\ref{fig:exp_setup}(a). $^{85}$Rb ground-state atoms are prepared in a magneto-optical trap (MOT) at a density of $\gtrsim 10^{10}$~cm$^{-3}$. The two-photon Rydberg excitation to $70S_{1/2}$ is driven by simultaneous 780~nm and 480~nm laser pulses with a 5~$\mu$s duration and $\approx$1~GHz red-detuning from the 5$P_{3/2}$ intermediate state.  Both beams propagate in the $xy$ plane and are linearly polarized along $\hat{\bf{z}}$.  The 780~nm beam has a Gaussian beam parameter $w_{0}$ of 0.75~mm and the 480~nm beam is focused to $w_{0} = 8~\mu$m.  The Rydberg atoms are ionized by applying a high voltage to a tip imaging probe (TIP) with a rounded tip of diameter 125~$\mu$m. Ions are accelerated by the TIP electric field towards a microchannel plate (MCP). Ion impacts result in blips produced by the MCP-phosphor detector assembly, revealing the Rydberg atom positions. In each experimental cycle we record one image, taken by a CCD camera; typically it contains several blips. For more detail see reference \citep{schwarzkopf_spatial_2013}.

The excitation volume is $\approx$470~$\mu$m above the tip, which, in combination with the radial divergence of the TIP electric field and the MCP front voltage, results in a measured magnification of 155 with an uncertainty of 2$\%$. To calibrate the magnification, we generate a Rydberg-atom distribution with a known spatial period by imaging the 480~nm beam through a double slit into the excitation region. The spatial period of the signal on the MCP then yields the magnification. The resolution in the object plane is $\approx 1~\mu$m; it follows from the magnification and the point spread function of the detector. We can easily resolve the correlation length between the Rydberg atoms, which is on the order of 10~$\mu$m. The field of view typically contains 5 to 10 detected atoms. In the image processing sequence, we first extract ion impact positions on the MCP by using a peak detection algorithm~\citep{schwarzkopf_spatial_2013}. Out of 10000 images taken in each dataset, we select the 5000 images with the highest numbers of detected ions. From this subset, we calculate the average pair correlation image and normalize it such that at large distances it approaches the value of one.

\begin{figure}[t]
	\centering				
			  \includegraphics[width=1\linewidth]{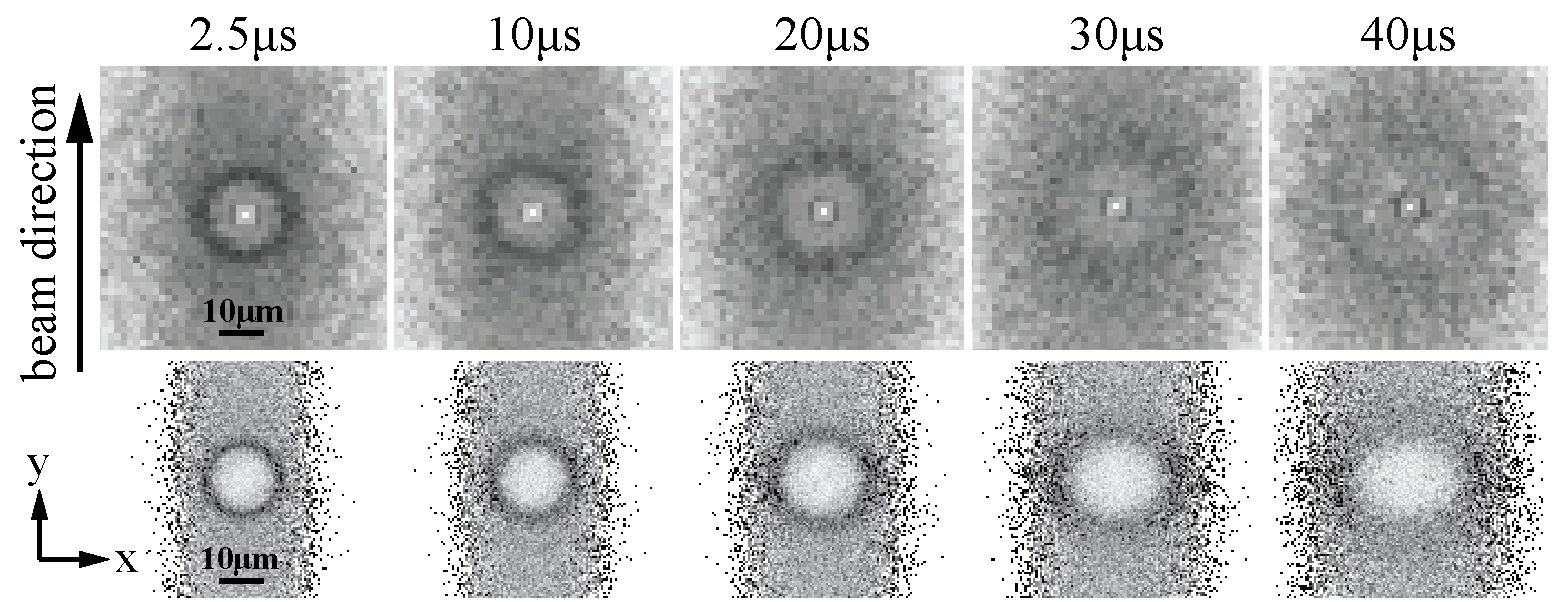}				
	\caption{Pair correlation images from experiment (top row) and simulation (bottom row) for the selected wait times. The linear grayscale ranges from 0 (white) to 2 (black). Values of 1, $<$1, and $>$1 indicate no correlation, anticorrelation, and positive correlation, respectively.}	
	\label{fig:trajectory_corr_exp}
\end{figure}		

The excitation of the $70S_{1/2}$ Rydberg level is detuned by $\delta\nu_{\rm{L}}=4\pm2$~MHz with respect to the two-photon resonance. The positive excitation detuning is compensated by the (repulsive) van der Waals interaction. We preferentially excite Rydberg atoms in pairs at a separation at which the van der Waals interaction is $2\delta\nu_{\rm{L}}=8\pm$4~MHz. We determine $r_{0}$ from pair-correlation images. After excitation, the Rydberg atoms are allowed to move within selected times before their positions are measured by applying a field ionization pulse [Fig.~\ref{fig:exp_setup}(b)]. The atom-position data are processed as explained above, yielding average pair correlation images. In Fig.~\ref{fig:trajectory_corr_exp}, we present the pair correlation images for wait times 2.5~$\mu$s, 10~$\mu$s, 20~$\mu$s, 30~$\mu$s, and 40~$\mu$s. The average interaction time of the Rydberg atoms is the wait time plus half the excitation pulse length (wait time + 2.5~$\mu$s). The pair correlation image at 2.5~$\mu$s exhibits strong correlation enhancement at a fairly well-defined radius. The initial correlation is critical for our trajectory experiment; it is sufficient to track Rydberg-pair trajectories out to approximately 40~$\mu$s. With increasing wait time, the radius of enhanced pair correlation increases, reflecting an increase of interatomic separation due to the repulsive van der Waals interaction. At long wait times, the enhancement ring is blurred out due to initial thermal atom velocities in the MOT (temperature $\sim100~\mu$K).

\begin{figure}[t]
\centering			
\includegraphics[width=1\linewidth]{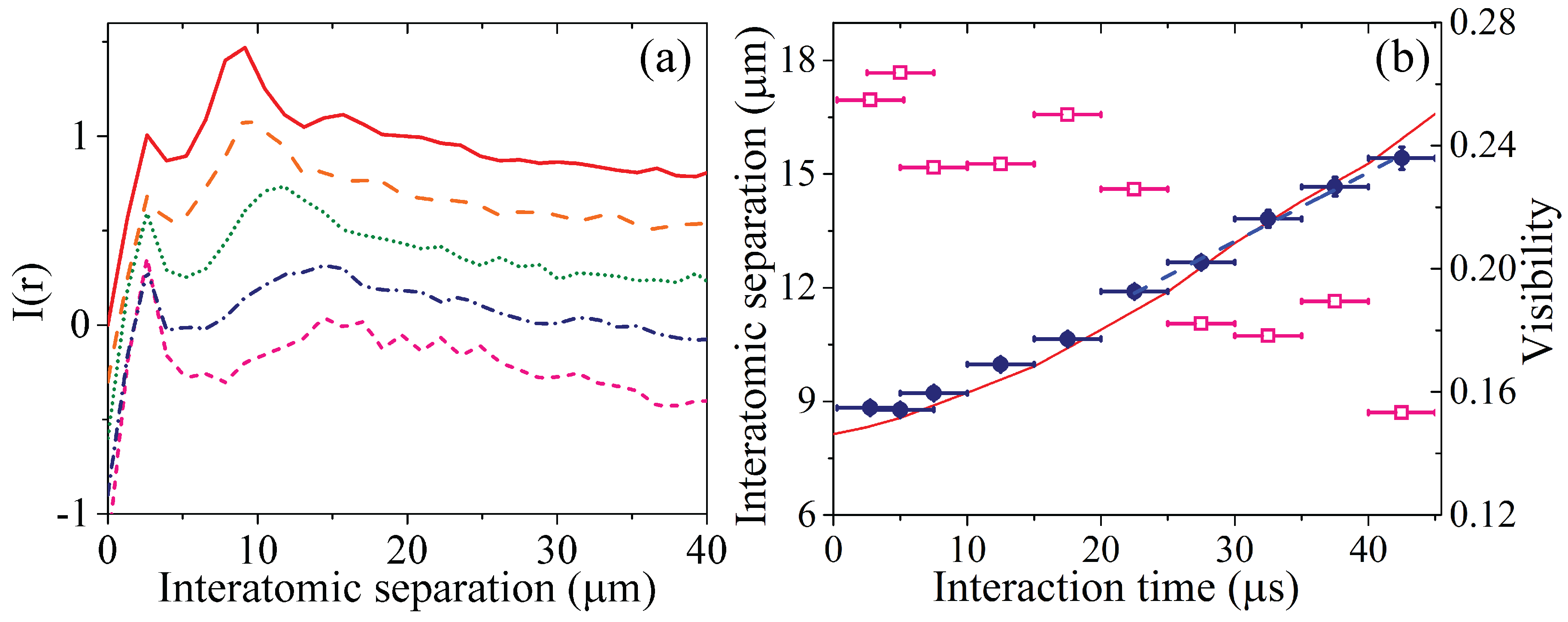}
\caption{(Color online) (a) Angular integrals $I(r)$ of the pair correlation images in Fig.~\ref{fig:trajectory_corr_exp} at wait times: 2.5~$\mu$s (red solid), 10~$\mu$s (orange dashed), 20~$\mu$s (green dotted), 30~$\mu$s (blue dot-dashed), and 40~$\mu$s (pink short-dashed). The $y$ axis is for the red-solid curve; for clarity, the other curves are shifted down in equidistant intervals of 0.3. (b) Interatomic separations between Rydberg-atom pairs as a function of interaction time (left axis, blue circles), obtained from the peak positions in the $I(r)$ curves. The blue dashed line represents a linear fit at long wait times (20-40~$\mu$s). The red solid line shows simulation results obtained for $\delta\nu_{\rm{L}}=3$~MHz. The pink hollow squares show the visibility of the experimental pair correlation enhancement (right axis) as defined in text.}	
\label{fig:Ir_All}
\end{figure}

To determine the most probable separation $r_{\rm{p}}(t)$ between Rydberg-atom pairs at wait time $t$, we first compute the angular integrals $I(r)$ of the experimental pair correlation images in Fig.~\ref{fig:trajectory_corr_exp}, as shown in Fig.~\ref{fig:Ir_All}(a). The values of $r_{\rm{p}}(t)$ are obtained from local parabolic fits centered approximately at the peak positions of the $I(r)$ curves. We include 4 to 7 data points in the fit (depending on the shapes of the curves). The resulting separations $r_{\rm{p}}(t)$ are shown in Fig.~\ref{fig:Ir_All}(b). The visibility of the pair correlation enhancement, also shown in Fig.~\ref{fig:Ir_All}(b), is $(I_{\rm{max}}-I_{\rm{min}})/(I_{\rm{max}}+I_{\rm{min}})$, where $I_{\rm{max}}$ is the peak value of $I(r)$ in the range $\gtrsim$ 10~$\mu$m and $I_{\rm{min}}$ is the minimal value near 7~$\mu$m. A cursory inspection of Fig.~\ref{fig:Ir_All}(b) already shows that the trajectory of the Rydberg-atom pairs is characterized by an initial acceleration phase, during which the initial van der Waals potential energy, $W_{0}$, is converted into kinetic energy, and a later phase during which the atoms keep separating at a fixed velocity. The drop in visibility is due to the thermally-induced blurring of the correlation ring at late times.

To extract the van der Waals $C_{6}$ coefficient, one may consider an isolated atom pair excited at an initial separation $r_{0}$. The initial van der Waals energy is
\begin{equation}
\label{eq:W0}
W_{0}=\frac{C_{6}}{r_{0}^{6}}=2\delta\nu_{\rm{L}}.
\end{equation}
From Eq.~\ref{eq:W0}, $C_{6}=2\delta\nu_{\rm{L}}r_{0}^{6}$ can, in principle, be obtained from spectroscopic measurement \citep{beguin_direct_2013}. This method requires a well-defined $r_{0}$, a narrow laser linewidth, and accurate knowledge of $\delta\nu_{\rm{L}}$. In our work, the relative uncertainty in $r_{0}^{6}$ is about 12\% (because the relative magnification uncertainty is 2\%). In comparison, the uncertainty arising from $\delta\nu_{\rm{L}}=4\pm2$~MHz is much larger. Taking all uncertainties into account, Eq.~\ref{eq:W0} leads to values of $C_{6}$ ranging from 7$\times10^{-58}$~Jm$^{6}$ to 40$\times10^{-58}$~Jm$^{6}$. Therefore, Eq.~\ref{eq:W0} only allows us to perform a crude order-of-magnitude estimate for $C_{6}$. The main weakness of Eq.~\ref{eq:W0} is that the $C_{6}$-values derived from it directly reflect our large relative uncertainty in $\delta\nu_{\rm{L}}$.

A better way for us to obtain $C_{6}$ is to track the evolution of the Rydberg-atom trajectories. Over the experimentally investigated wait times, the entire initial van der Waals energy $W_{0}$ becomes converted into kinetic energy. With the reduced mass of the atom pair, $\mu$, and the terminal relative velocity, $v_{\rm{t}}$, it is
\begin{equation}
\label{eq:energy_conservation}
 W_{0}=\frac{C_{6}}{r_{0}^{6}}=\frac{1}{2}\mu v_{\rm{t}}^{2},
\end{equation}
To obtain $r_{0}$, we take the weighted average of the fit results $r_{\rm{p}}(t)$ at the earliest times used in the experiment [the points in Fig.~\ref{fig:Ir_All}(b) at 250~ns and 2.5~$\mu$s]. The averaging is valid because during the first few microseconds after excitation the Rydberg atoms are frozen in place due to their inertia. The statistical weights are given by the inverse squares of the fitting uncertainties of $r_{\rm{p}}$ at 250~ns and at 2.5~$\mu$s.  We obtain $r_{0}=8.78~\mu$m with a net fitting uncertainty of 0.02~$\mu$m. To determine $v_{\rm{t}}$, we perform a linear fit at long wait times (20-40~$\mu$s) and obtain $v_{\rm{t}}=0.182$~m/s with a fitting uncertainty of 0.008~m/s. It follows $C_{6}=\frac{1}{2}\mu v_{\rm{t}}^{2}r_{0}^{6}=(5.4\pm0.5)\times10^{-58}$~Jm$^{6}$. Including the 2\% magnification uncertainty, the total uncertainty of $C_{6}$ becomes 1.0~$\times10^{-58}$~Jm$^{6}$.

The final relative uncertainty of $C_{6}$ follows from three statistically independent contributions: the magnification uncertainty, the fit uncertainty for $v_{\rm{t}}$, and the fit uncertainty for $r_{0}$. The respective powers at which these quantities enter into $C_{6}$ are 8, 2, and 6. Factoring in these powers, the three quantities contribute respective independent uncertainties of 16\%, 8.9\%, and 1.4\% to the relative uncertainty of $C_{6}$, leading to the total relative uncertainty of 18\%. In our method, achieving a small magnification uncertainty  is particularly important.

A semi-classical 3D simulation of the dynamics of Rydberg atoms interacting by an isotropic van der Waals force has been performed to interpret the experiment. The simulation volume of (140~$\mu$m)$^{3}$ exceeds the experimentally relevant volume by about a factor of two in each dimension. The 480~nm excitation beam propagates along the $y$ direction. The excitation volume in the transverse directions ($x$ and $z$) is limited by the size of the excitation beam ($w_{0}=8~\mu$m). The number of simulations is 5000, in keeping with the number of images analyzed in the experiment. Rydberg-atom positions and velocities are initialized as explained below, and the positions and velocities are then propagated using a Runge-Kutta integrator that includes all pair-wise interatomic forces. We use $C_{6}=5.7\times10^{-58}$Jm$^{6}$ \citep{reinhard_level_2007} in the initialization and integration procedures. To avoid edge effects, we crop the $xy$ processing area in the images to (70~$\mu$m)$^{2}$ before calculating the average pair-correlation images and the radial functions $I(r)$.

In each simulation, we start by drawing the number of Rydberg atoms to be used, $n_{\rm{Ryd}}$, from a Poissonian distribution. Initial trial positions of the Rydberg atoms are then drawn from a probability distribution that is Gaussian along $x$ and $z$, with a $w_{0}$ of 8~$\mu$m, and uniform along $y$ (in close analogy with the experiment). For an atom $i$ at a trial position ${\bf{r}}_{i}$, the effective detuning $\delta_{\rm{eff}}({\bf{r}}_{i})$ due to all other atoms $j$ at positions ${\bf{r}}_{j}$ which are already excited into the Rydberg state is
\begin{equation}
\label{eq:effective detuning}
\delta_{\rm{eff}}({\bf{r}}_{i}) = \sum\limits_{j}\frac{C_{6}}{|{\bf{r}}_{i}-{\bf{r}}_{j}|^{6}}-\delta.
\end{equation}
Since the detuning $\delta\nu_{\rm{L}}$ is substantial, the first pair of Rydberg atoms is simultaneously excited via off-resonant excitation \citep{schwarzkopf_spatial_2013} (because the intermediate state in which there is only one Rydberg atom present is off-resonant). Therefore the value of $\delta$ in Eq.~\ref{eq:effective detuning} for the first Rydberg-atom pair is set to be twice the laser detuning $\delta\nu_{\rm{L}}$. For the excitation of additional  atoms we use $\delta=\delta\nu_{\rm{L}}$ to simulate the stepwise addition of those atoms, which can be a near-resonant process with other atoms already present \citep{urvoy_strongly_2015}. Detunings due to the Doppler effect are about 300~kHz and are neglected. The excitation probability $P_{\rm{ex}}(\delta_{\rm{eff}})$ is a Gaussian centered at $\delta_{\rm{eff}}=0$ with a FWHM of 4~MHz (given by the excitation bandwidth) and $P_{\rm{ex}}(\delta_{\rm{eff}}=0)=1$. A new Rydberg atom at position ${\bf{r}}_{i}$ is created if the excitation probability is larger than a number randomly drawn between 0 and 1. This procedure is repeated with new trial positions until the desired number of $n_{\rm{Ryd}}$ Rydberg atoms has been reached. The initial center-of-mass velocities of the atoms are assigned using a Maxwell distribution at temperature 100~$\mu$K.

In the simulation, we record the atom positions at the same wait times as used in the experiment. The resulting pair correlation images are shown for the case $\delta\nu_{\rm{L}}=3$~MHz in the bottom row in Fig.~\ref{fig:trajectory_corr_exp}. Black pixels along the left and right edges of each pair correlation are an artifact due to the normalization used in the image processing.

From the simulated pair correlation images we calculate the $I(r)$ curves, as in the experiment, in order to obtain the most probable separations $r_{\rm{p}}(t)$ between Rydberg-atom pairs at each interaction time [see solid curve in Fig.~\ref{fig:Ir_All}(b), which is for $\delta\nu_{\rm{L}}=3$~MHz]. The simulated and experimental results for $r_{\rm{p}}(t)$ are in good agreement. In order to test how well our experimental procedure reproduces the $C_{6}$ coefficient that underlies the atomic interactions, we evaluate the simulated results for $r_{\rm{p}}(t)$ using the same method that we also employ for the experimental data. The results for $r_{0}$, $v_{\rm{t}}$ and $C_{6}$ extracted from the simulated $r_{\rm{p}}(t)$ curves are shown in Table~\ref{table:C6} for four choices of $\delta\nu_{\rm{L}}$.  These $C_{6}$ values agree, within the uncertainties, with the value that has been entered as a fixed input into the simulation. This finding validates the experimentally used procedure; in particular it is seen that the method is not very sensitive on $\delta\nu_{\rm{L}}$. The experimental and calculated~\cite{reinhard_level_2007} $C_{6}$ values are also included in Table~\ref{table:C6} for reference.

\begin{table}[htb]
    \caption{Comparison of calculated, experimental, and simulated results for $C_{6}$, $r_{0}$, and $v_{\rm{t}}$.}
    \centering{
	\setlength{\tabcolsep}{4pt}
    \begin{tabular}{c c c c} 
    \hline \hline 		
	~ & $r_{0}$& $v_{\rm{t}}$& $C_{6}$ \\			
	~ & ($\mu$m) & (m/s) & ($\times10^{-58}$Jm$^{6}$) \\		
	\hline	
	Calculation	& ~ & ~ & 5.7 \\ 
	Experiment	& 8.8$\pm$0.2 &	0.182$\pm$0.009 & 5.4$\pm$1.0 \\
	Simulations & ~ & ~ & ~ \\
	$\delta\nu_{\rm{L}}=2$~MHz & 8.85$\pm$0.06 & 0.17$\pm$0.02 & 4.9 $\pm$1.0 \\
    $\delta\nu_{\rm{L}}=3$~MHz & 8.14$\pm$0.03 & 0.233$\pm$0.005 & 5.5$\pm$0.3 \\		
    $\delta\nu_{\rm{L}}=4$~MHz & 7.76$\pm$0.05 & 0.278$\pm$0.003 & 6.0$\pm$0.3 \\
    $\delta\nu_{\rm{L}}=5$~MHz & 7.51$\pm$0.04 & 0.28$\pm$0.01 & 4.9$\pm$0.5 \\ 
    \hline 
    \end{tabular}}
    \label{table:C6}
\end{table}

The experimental, simulated, and expected values for $C_{6}$ in Table~\ref{table:C6} are in reasonable agreement. Moreover, we observe that the entire simulated curve $r_{\rm{p}}(t)$ for $\delta\nu_{\rm{L}}=3$~MHz matches the experimental result very well [see Figs.~\ref{fig:trajectory_corr_exp} and~\ref{fig:Ir_All}(b)]. Overall, our findings lend credibility to our method of measuring $C_{6}$.

We note that higher-order quadrupole-dipole and quadrupole-quadrupole interaction potentials, which scale as $r^{-7}$ and $r^{-8}$, are not important at the distances relevant in our work, at the current level of precision. We have verified this in a calculation of interaction potentials in which all terms up to the quadrupole-quadrupole interaction have been included~\citep{supplement}.

A question of interest is whether the interatomic force is acting on ``superatoms'', {\sl {i.e.}} Rydberg excitations shared among a number of ground-state atoms, or on individual atoms. In our case, there are several tens of atoms within one blockade sphere. In our analysis, we have assumed that the effective mass of the interacting entities is half the rubidium atom mass, {\sl {i.e.}} we have implied that the interacting entities are individual atoms and not superatoms. For our density and blockade radius, the total mass of a superatom equals that of several tens of atoms. The agreement between the $C_{6}$ coefficients in Table~\ref{table:C6} demonstrates that the interacting entities are indeed individual atoms. The finding implies that, during the course of the van der Waals interaction, excitations within superatoms become projected onto individual atoms, which are then ejected from the initial superatom volumes. The phenomenon has been predicted in \citep{mobius_breakup_2013} for superatom clouds interacting via a dipole-dipole interaction. In our work, we arrive at a similar conclusion for van-der-Waals-interacting Rydberg atoms in a 3D system.

\begin{figure}[t]
\centering	
\includegraphics[width=0.8\linewidth]{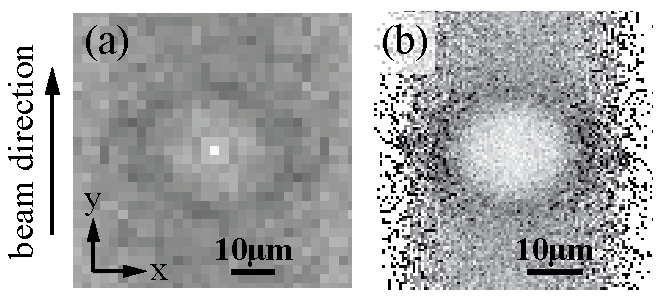}
\caption{Pair correlation function at wait time 30 $\mu$s from (a) experiment and (b) simulation at $\delta\nu_{\rm{L}}=3$~MHz. The anisotropic expansion of the atom ensemble causes the radius of enhanced pair correlation along the transverse direction ($x$) to be larger than along the excitation beam direction ($y$).}	
\label{fig:anisotropy}
\end{figure}

Close inspection of the pair correlation functions from the simulation presented in Fig.~\ref{fig:trajectory_corr_exp} reveals  anisotropic expansion behavior: the radius of enhanced pair correlation is larger in $x$ than in $y$ direction. In several experimental data sets we see some evidence of anisotropic expansion, such as in the experimental result shown in Fig.~\ref{fig:anisotropy}(a). The anisotropic expansion is due to the cylindrical shape of the excitation volume (not the interatomic interaction, which is isotropic). Since the blockade radius is close to the transverse size of the excitation region, at most two atoms can be excited side-by-side in $x$ direction, leading to free, unimpeded expansion along that direction. In contrast, more than two Rydberg atoms can be created along the $y$ direction. Therefore, along $y$ the expansion is slowed down due to multiple-atom repulsion. In our experiment, the overall Rydberg-atom density is high enough to sometimes observe this effect.

In summary, we have studied the trajectory of Rydberg-atom pairs interacting by repulsive, isotropic van der Waals interaction. We have extracted the $C_{6}$ coefficient from the experiment and compared it with simulations and calculations, and have observed good agreement. The result implies that the interaction occurs between individual atoms, not superatoms. We have observed indications of an anisotropic effect in the expansion, caused by the excitation geometry. Future work may involve atom-atom interactions that are anisotropic, such as the dipole-dipole interaction.

This work was supported by the NSF (PHY-1205559) and the AFOSR (FA9550-10-1-0453).  NT acknowledges support from DPST of Thailand.

\noindent \\
*nithi@umich.edu\\
$^{\dag}$Present address: zeroK NanoTech Corporation, Gaithersburg, MD 20878, USA
\newpage

\bibstyle{apsrev4-1}

\begin{thebibliography}{24}%
\makeatletter
\providecommand \@ifxundefined [1]{%
 \@ifx{#1\undefined}
}%
\providecommand \@ifnum [1]{%
 \ifnum #1\expandafter \@firstoftwo
 \else \expandafter \@secondoftwo
 \fi
}%
\providecommand \@ifx [1]{%
 \ifx #1\expandafter \@firstoftwo
 \else \expandafter \@secondoftwo
 \fi
}%
\providecommand \natexlab [1]{#1}%
\providecommand \enquote  [1]{``#1''}%
\providecommand \bibnamefont  [1]{#1}%
\providecommand \bibfnamefont [1]{#1}%
\providecommand \citenamefont [1]{#1}%
\providecommand \href@noop [0]{\@secondoftwo}%
\providecommand \href [0]{\begingroup \@sanitize@url \@href}%
\providecommand \@href[1]{\@@startlink{#1}\@@href}%
\providecommand \@@href[1]{\endgroup#1\@@endlink}%
\providecommand \@sanitize@url [0]{\catcode `\\12\catcode `\$12\catcode
  `\&12\catcode `\#12\catcode `\^12\catcode `\_12\catcode `\%12\relax}%
\providecommand \@@startlink[1]{}%
\providecommand \@@endlink[0]{}%
\providecommand \url  [0]{\begingroup\@sanitize@url \@url }%
\providecommand \@url [1]{\endgroup\@href {#1}{\urlprefix }}%
\providecommand \urlprefix  [0]{URL }%
\providecommand \Eprint [0]{\href }%
\providecommand \doibase [0]{http://dx.doi.org/}%
\providecommand \selectlanguage [0]{\@gobble}%
\providecommand \bibinfo  [0]{\@secondoftwo}%
\providecommand \bibfield  [0]{\@secondoftwo}%
\providecommand \translation [1]{[#1]}%
\providecommand \BibitemOpen [0]{}%
\providecommand \bibitemStop [0]{}%
\providecommand \bibitemNoStop [0]{.\EOS\space}%
\providecommand \EOS [0]{\spacefactor3000\relax}%
\providecommand \BibitemShut  [1]{\csname bibitem#1\endcsname}%
\let\auto@bib@innerbib\@empty
\bibitem [{\citenamefont {Lukin}\ \emph {et~al.}(2001)\citenamefont {Lukin},
  \citenamefont {Fleischhauer}, \citenamefont {Cote}, \citenamefont {Duan},
  \citenamefont {Jaksch}, \citenamefont {Cirac},\ and\ \citenamefont
  {Zoller}}]{lukin_dipole_2001}%
  \BibitemOpen
  \bibfield  {author} {\bibinfo {author} {\bibfnamefont {M.~D.}\ \bibnamefont
  {Lukin}}, \bibinfo {author} {\bibfnamefont {M.}~\bibnamefont {Fleischhauer}},
  \bibinfo {author} {\bibfnamefont {R.}~\bibnamefont {Cote}}, \bibinfo {author}
  {\bibfnamefont {L.~M.}\ \bibnamefont {Duan}}, \bibinfo {author}
  {\bibfnamefont {D.}~\bibnamefont {Jaksch}}, \bibinfo {author} {\bibfnamefont
  {J.~I.}\ \bibnamefont {Cirac}}, \ and\ \bibinfo {author} {\bibfnamefont
  {P.}~\bibnamefont {Zoller}},\ }\href {\doibase 10.1103/PhysRevLett.87.037901}
  {\bibfield  {journal} {\bibinfo  {journal} {Phys. Rev. Lett.}\ }\textbf
  {\bibinfo {volume} {87}},\ \bibinfo {pages} {037901} (\bibinfo {year}
  {2001})}\BibitemShut {NoStop}%
\bibitem [{\citenamefont {Tong}\ \emph {et~al.}(2004)\citenamefont {Tong},
  \citenamefont {Farooqi}, \citenamefont {Stanojevic}, \citenamefont
  {Krishnan}, \citenamefont {Zhang}, \citenamefont {Côté}, \citenamefont
  {Eyler},\ and\ \citenamefont {Gould}}]{tong_local_2004}%
  \BibitemOpen
  \bibfield  {author} {\bibinfo {author} {\bibfnamefont {D.}~\bibnamefont
  {Tong}}, \bibinfo {author} {\bibfnamefont {S.~M.}\ \bibnamefont {Farooqi}},
  \bibinfo {author} {\bibfnamefont {J.}~\bibnamefont {Stanojevic}}, \bibinfo
  {author} {\bibfnamefont {S.}~\bibnamefont {Krishnan}}, \bibinfo {author}
  {\bibfnamefont {Y.~P.}\ \bibnamefont {Zhang}}, \bibinfo {author}
  {\bibfnamefont {R.}~\bibnamefont {Côté}}, \bibinfo {author} {\bibfnamefont
  {E.~E.}\ \bibnamefont {Eyler}}, \ and\ \bibinfo {author} {\bibfnamefont
  {P.~L.}\ \bibnamefont {Gould}},\ }\href {\doibase
  10.1103/PhysRevLett.93.063001} {\bibfield  {journal} {\bibinfo  {journal}
  {Phys. Rev. Lett.}\ }\textbf {\bibinfo {volume} {93}},\ \bibinfo {pages}
  {063001} (\bibinfo {year} {2004})}\BibitemShut {NoStop}%
\bibitem [{\citenamefont {Liebisch}\ \emph {et~al.}(2005)\citenamefont
  {Liebisch}, \citenamefont {Reinhard}, \citenamefont {Berman},\ and\
  \citenamefont {Raithel}}]{liebisch_atom_2005}%
  \BibitemOpen
  \bibfield  {author} {\bibinfo {author} {\bibfnamefont {T.~C.}\ \bibnamefont
  {Liebisch}}, \bibinfo {author} {\bibfnamefont {A.}~\bibnamefont {Reinhard}},
  \bibinfo {author} {\bibfnamefont {P.~R.}\ \bibnamefont {Berman}}, \ and\
  \bibinfo {author} {\bibfnamefont {G.}~\bibnamefont {Raithel}},\ }\href
  {\doibase 10.1103/PhysRevLett.95.253002} {\bibfield  {journal} {\bibinfo
  {journal} {Phys. Rev. Lett.}\ }\textbf {\bibinfo {volume} {95}},\ \bibinfo
  {pages} {253002} (\bibinfo {year} {2005})}\BibitemShut {NoStop}%
\bibitem [{\citenamefont {Gärttner}\ \emph {et~al.}(2014)\citenamefont
  {Gärttner}, \citenamefont {Whitlock}, \citenamefont {Schönleber},\ and\
  \citenamefont {Evers}}]{garttner_collective_2014}%
  \BibitemOpen
  \bibfield  {author} {\bibinfo {author} {\bibfnamefont {M.}~\bibnamefont
  {Gärttner}}, \bibinfo {author} {\bibfnamefont {S.}~\bibnamefont {Whitlock}},
  \bibinfo {author} {\bibfnamefont {D.~W.}\ \bibnamefont {Schönleber}}, \ and\
  \bibinfo {author} {\bibfnamefont {J.}~\bibnamefont {Evers}},\ }\href
  {\doibase 10.1103/PhysRevLett.113.233002} {\bibfield  {journal} {\bibinfo
  {journal} {Phys. Rev. Lett.}\ }\textbf {\bibinfo {volume} {113}},\ \bibinfo
  {pages} {233002} (\bibinfo {year} {2014})}\BibitemShut {NoStop}%
\bibitem [{\citenamefont {Pohl}\ \emph {et~al.}(2010)\citenamefont {Pohl},
  \citenamefont {Demler},\ and\ \citenamefont {Lukin}}]{pohl_dynamical_2010}%
  \BibitemOpen
  \bibfield  {author} {\bibinfo {author} {\bibfnamefont {T.}~\bibnamefont
  {Pohl}}, \bibinfo {author} {\bibfnamefont {E.}~\bibnamefont {Demler}}, \ and\
  \bibinfo {author} {\bibfnamefont {M.~D.}\ \bibnamefont {Lukin}},\ }\href
  {\doibase 10.1103/PhysRevLett.104.043002} {\bibfield  {journal} {\bibinfo
  {journal} {Phys. Rev. Lett.}\ }\textbf {\bibinfo {volume} {104}},\ \bibinfo
  {pages} {043002} (\bibinfo {year} {2010})}\BibitemShut {NoStop}%
\bibitem [{\citenamefont {Schauß}\ \emph {et~al.}(2014)\citenamefont
  {Schauß}, \citenamefont {Zeiher}, \citenamefont {Fukuhara}, \citenamefont
  {Hild}, \citenamefont {Cheneau}, \citenamefont {Macrì}, \citenamefont
  {Pohl}, \citenamefont {Bloch},\ and\ \citenamefont
  {Gross}}]{schaus_dynamical_2014}%
  \BibitemOpen
  \bibfield  {author} {\bibinfo {author} {\bibfnamefont {P.}~\bibnamefont
  {Schauß}}, \bibinfo {author} {\bibfnamefont {J.}~\bibnamefont {Zeiher}},
  \bibinfo {author} {\bibfnamefont {T.}~\bibnamefont {Fukuhara}}, \bibinfo
  {author} {\bibfnamefont {S.}~\bibnamefont {Hild}}, \bibinfo {author}
  {\bibfnamefont {M.}~\bibnamefont {Cheneau}}, \bibinfo {author} {\bibfnamefont
  {T.}~\bibnamefont {Macrì}}, \bibinfo {author} {\bibfnamefont
  {T.}~\bibnamefont {Pohl}}, \bibinfo {author} {\bibfnamefont {I.}~\bibnamefont
  {Bloch}}, \ and\ \bibinfo {author} {\bibfnamefont {C.}~\bibnamefont
  {Gross}},\ }\href {http://arxiv.org/abs/1404.0980} {\bibfield  {journal}
  {\bibinfo  {journal} {arXiv:1404.0980 [cond-mat, physics:physics]}\ }
  (\bibinfo {year} {2014})},\ \bibinfo {note} {arXiv: 1404.0980}\BibitemShut
  {NoStop}%
\bibitem [{\citenamefont {Lesanovsky}\ and\ \citenamefont
  {Garrahan}(2014)}]{lesanovsky_out--equilibrium_2014}%
  \BibitemOpen
  \bibfield  {author} {\bibinfo {author} {\bibfnamefont {I.}~\bibnamefont
  {Lesanovsky}}\ and\ \bibinfo {author} {\bibfnamefont {J.~P.}\ \bibnamefont
  {Garrahan}},\ }\href {\doibase 10.1103/PhysRevA.90.011603} {\bibfield
  {journal} {\bibinfo  {journal} {Phys. Rev. A}\ }\textbf {\bibinfo {volume}
  {90}},\ \bibinfo {pages} {011603} (\bibinfo {year} {2014})}\BibitemShut
  {NoStop}%
\bibitem [{\citenamefont {Urvoy}\ \emph {et~al.}(2015)\citenamefont {Urvoy},
  \citenamefont {Ripka}, \citenamefont {Lesanovsky}, \citenamefont {Booth},
  \citenamefont {Shaffer}, \citenamefont {Pfau},\ and\ \citenamefont
  {Löw}}]{urvoy_strongly_2015}%
  \BibitemOpen
  \bibfield  {author} {\bibinfo {author} {\bibfnamefont {A.}~\bibnamefont
  {Urvoy}}, \bibinfo {author} {\bibfnamefont {F.}~\bibnamefont {Ripka}},
  \bibinfo {author} {\bibfnamefont {I.}~\bibnamefont {Lesanovsky}}, \bibinfo
  {author} {\bibfnamefont {D.}~\bibnamefont {Booth}}, \bibinfo {author}
  {\bibfnamefont {J.}~\bibnamefont {Shaffer}}, \bibinfo {author} {\bibfnamefont
  {T.}~\bibnamefont {Pfau}}, \ and\ \bibinfo {author} {\bibfnamefont
  {R.}~\bibnamefont {Löw}},\ }\href {\doibase 10.1103/PhysRevLett.114.203002}
  {\bibfield  {journal} {\bibinfo  {journal} {Phys. Rev. Lett.}\ }\textbf
  {\bibinfo {volume} {114}},\ \bibinfo {pages} {203002} (\bibinfo {year}
  {2015})}\BibitemShut {NoStop}%
\bibitem [{\citenamefont {Isenhower}\ \emph {et~al.}(2010)\citenamefont
  {Isenhower}, \citenamefont {Urban}, \citenamefont {Zhang}, \citenamefont
  {Gill}, \citenamefont {Henage}, \citenamefont {Johnson}, \citenamefont
  {Walker},\ and\ \citenamefont {Saffman}}]{isenhower_demonstration_2010}%
  \BibitemOpen
  \bibfield  {author} {\bibinfo {author} {\bibfnamefont {L.}~\bibnamefont
  {Isenhower}}, \bibinfo {author} {\bibfnamefont {E.}~\bibnamefont {Urban}},
  \bibinfo {author} {\bibfnamefont {X.~L.}\ \bibnamefont {Zhang}}, \bibinfo
  {author} {\bibfnamefont {A.~T.}\ \bibnamefont {Gill}}, \bibinfo {author}
  {\bibfnamefont {T.}~\bibnamefont {Henage}}, \bibinfo {author} {\bibfnamefont
  {T.~A.}\ \bibnamefont {Johnson}}, \bibinfo {author} {\bibfnamefont {T.~G.}\
  \bibnamefont {Walker}}, \ and\ \bibinfo {author} {\bibfnamefont
  {M.}~\bibnamefont {Saffman}},\ }\href {\doibase
  10.1103/PhysRevLett.104.010503} {\bibfield  {journal} {\bibinfo  {journal}
  {Phys. Rev. Lett.}\ }\textbf {\bibinfo {volume} {104}},\ \bibinfo {pages}
  {010503} (\bibinfo {year} {2010})}\BibitemShut {NoStop}%
\bibitem [{\citenamefont {Wilk}\ \emph {et~al.}(2010)\citenamefont {Wilk},
  \citenamefont {Gaëtan}, \citenamefont {Evellin}, \citenamefont {Wolters},
  \citenamefont {Miroshnychenko}, \citenamefont {Grangier},\ and\ \citenamefont
  {Browaeys}}]{wilk_entanglement_2010}%
  \BibitemOpen
  \bibfield  {author} {\bibinfo {author} {\bibfnamefont {T.}~\bibnamefont
  {Wilk}}, \bibinfo {author} {\bibfnamefont {A.}~\bibnamefont {Gaëtan}},
  \bibinfo {author} {\bibfnamefont {C.}~\bibnamefont {Evellin}}, \bibinfo
  {author} {\bibfnamefont {J.}~\bibnamefont {Wolters}}, \bibinfo {author}
  {\bibfnamefont {Y.}~\bibnamefont {Miroshnychenko}}, \bibinfo {author}
  {\bibfnamefont {P.}~\bibnamefont {Grangier}}, \ and\ \bibinfo {author}
  {\bibfnamefont {A.}~\bibnamefont {Browaeys}},\ }\href {\doibase
  10.1103/PhysRevLett.104.010502} {\bibfield  {journal} {\bibinfo  {journal}
  {Phys. Rev. Lett.}\ }\textbf {\bibinfo {volume} {104}},\ \bibinfo {pages}
  {010502} (\bibinfo {year} {2010})}\BibitemShut {NoStop}%
\bibitem [{\citenamefont {Keating}\ \emph {et~al.}(2013)\citenamefont
  {Keating}, \citenamefont {Goyal}, \citenamefont {Jau}, \citenamefont
  {Biedermann}, \citenamefont {Landahl},\ and\ \citenamefont
  {Deutsch}}]{keating_adiabatic_2013}%
  \BibitemOpen
  \bibfield  {author} {\bibinfo {author} {\bibfnamefont {T.}~\bibnamefont
  {Keating}}, \bibinfo {author} {\bibfnamefont {K.}~\bibnamefont {Goyal}},
  \bibinfo {author} {\bibfnamefont {Y.-Y.}\ \bibnamefont {Jau}}, \bibinfo
  {author} {\bibfnamefont {G.~W.}\ \bibnamefont {Biedermann}}, \bibinfo
  {author} {\bibfnamefont {A.~J.}\ \bibnamefont {Landahl}}, \ and\ \bibinfo
  {author} {\bibfnamefont {I.~H.}\ \bibnamefont {Deutsch}},\ }\href {\doibase
  10.1103/PhysRevA.87.052314} {\bibfield  {journal} {\bibinfo  {journal} {Phys.
  Rev. A}\ }\textbf {\bibinfo {volume} {87}},\ \bibinfo {pages} {052314}
  (\bibinfo {year} {2013})}\BibitemShut {NoStop}%
\bibitem [{\citenamefont {Petrosyan}\ and\ \citenamefont
  {Mølmer}(2014)}]{petrosyan_binding_2014}%
  \BibitemOpen
  \bibfield  {author} {\bibinfo {author} {\bibfnamefont {D.}~\bibnamefont
  {Petrosyan}}\ and\ \bibinfo {author} {\bibfnamefont {K.}~\bibnamefont
  {Mølmer}},\ }\href {\doibase 10.1103/PhysRevLett.113.123003} {\bibfield
  {journal} {\bibinfo  {journal} {Phys. Rev. Lett.}\ }\textbf {\bibinfo
  {volume} {113}},\ \bibinfo {pages} {123003} (\bibinfo {year}
  {2014})}\BibitemShut {NoStop}%
\bibitem [{\citenamefont {Reinhard}\ \emph {et~al.}(2008)\citenamefont
  {Reinhard}, \citenamefont {Younge}, \citenamefont {Liebisch}, \citenamefont
  {Knuffman}, \citenamefont {Berman},\ and\ \citenamefont
  {Raithel}}]{reinhard_double-resonance_2008}%
  \BibitemOpen
  \bibfield  {author} {\bibinfo {author} {\bibfnamefont {A.}~\bibnamefont
  {Reinhard}}, \bibinfo {author} {\bibfnamefont {K.~C.}\ \bibnamefont
  {Younge}}, \bibinfo {author} {\bibfnamefont {T.~C.}\ \bibnamefont
  {Liebisch}}, \bibinfo {author} {\bibfnamefont {B.}~\bibnamefont {Knuffman}},
  \bibinfo {author} {\bibfnamefont {P.~R.}\ \bibnamefont {Berman}}, \ and\
  \bibinfo {author} {\bibfnamefont {G.}~\bibnamefont {Raithel}},\ }\href
  {\doibase 10.1103/PhysRevLett.100.233201} {\bibfield  {journal} {\bibinfo
  {journal} {Phys. Rev. Lett.}\ }\textbf {\bibinfo {volume} {100}},\ \bibinfo
  {pages} {233201} (\bibinfo {year} {2008})}\BibitemShut {NoStop}%
\bibitem [{\citenamefont {Béguin}\ \emph {et~al.}(2013)\citenamefont
  {Béguin}, \citenamefont {Vernier}, \citenamefont {Chicireanu}, \citenamefont
  {Lahaye},\ and\ \citenamefont {Browaeys}}]{beguin_direct_2013}%
  \BibitemOpen
  \bibfield  {author} {\bibinfo {author} {\bibfnamefont {L.}~\bibnamefont
  {Béguin}}, \bibinfo {author} {\bibfnamefont {A.}~\bibnamefont {Vernier}},
  \bibinfo {author} {\bibfnamefont {R.}~\bibnamefont {Chicireanu}}, \bibinfo
  {author} {\bibfnamefont {T.}~\bibnamefont {Lahaye}}, \ and\ \bibinfo {author}
  {\bibfnamefont {A.}~\bibnamefont {Browaeys}},\ }\href {\doibase
  10.1103/PhysRevLett.110.263201} {\bibfield  {journal} {\bibinfo  {journal}
  {Phys. Rev. Lett.}\ }\textbf {\bibinfo {volume} {110}},\ \bibinfo {pages}
  {263201} (\bibinfo {year} {2013})}\BibitemShut {NoStop}%
\bibitem [{\citenamefont {Sandoghdar}\ \emph {et~al.}(1992)\citenamefont
  {Sandoghdar}, \citenamefont {Sukenik}, \citenamefont {Hinds},\ and\
  \citenamefont {Haroche}}]{sandoghdar_direct_1992}%
  \BibitemOpen
  \bibfield  {author} {\bibinfo {author} {\bibfnamefont {V.}~\bibnamefont
  {Sandoghdar}}, \bibinfo {author} {\bibfnamefont {C.~I.}\ \bibnamefont
  {Sukenik}}, \bibinfo {author} {\bibfnamefont {E.~A.}\ \bibnamefont {Hinds}},
  \ and\ \bibinfo {author} {\bibfnamefont {S.}~\bibnamefont {Haroche}},\ }\href
  {\doibase 10.1103/PhysRevLett.68.3432} {\bibfield  {journal} {\bibinfo
  {journal} {Phys. Rev. Lett.}\ }\textbf {\bibinfo {volume} {68}},\ \bibinfo
  {pages} {3432} (\bibinfo {year} {1992})}\BibitemShut {NoStop}%
\bibitem [{\citenamefont {Nordlander}\ and\ \citenamefont
  {Dunning}(1996)}]{nordlander_interaction_1996}%
  \BibitemOpen
  \bibfield  {author} {\bibinfo {author} {\bibfnamefont {P.}~\bibnamefont
  {Nordlander}}\ and\ \bibinfo {author} {\bibfnamefont {F.~B.}\ \bibnamefont
  {Dunning}},\ }\href {\doibase 10.1103/PhysRevB.53.8083} {\bibfield  {journal}
  {\bibinfo  {journal} {Phys. Rev. B}\ }\textbf {\bibinfo {volume} {53}},\
  \bibinfo {pages} {8083} (\bibinfo {year} {1996})}\BibitemShut {NoStop}%
\bibitem [{\citenamefont {Fichet}\ \emph {et~al.}(2007)\citenamefont {Fichet},
  \citenamefont {Dutier}, \citenamefont {Yarovitsky}, \citenamefont {Todorov},
  \citenamefont {Hamdi}, \citenamefont {Maurin}, \citenamefont {Saltiel},
  \citenamefont {Sarkisyan}, \citenamefont {Gorza}, \citenamefont {Bloch},\
  and\ \citenamefont {Ducloy}}]{fichet_exploring_2007}%
  \BibitemOpen
  \bibfield  {author} {\bibinfo {author} {\bibfnamefont {M.}~\bibnamefont
  {Fichet}}, \bibinfo {author} {\bibfnamefont {G.}~\bibnamefont {Dutier}},
  \bibinfo {author} {\bibfnamefont {A.}~\bibnamefont {Yarovitsky}}, \bibinfo
  {author} {\bibfnamefont {P.}~\bibnamefont {Todorov}}, \bibinfo {author}
  {\bibfnamefont {I.}~\bibnamefont {Hamdi}}, \bibinfo {author} {\bibfnamefont
  {I.}~\bibnamefont {Maurin}}, \bibinfo {author} {\bibfnamefont
  {S.}~\bibnamefont {Saltiel}}, \bibinfo {author} {\bibfnamefont
  {D.}~\bibnamefont {Sarkisyan}}, \bibinfo {author} {\bibfnamefont {M.-P.}\
  \bibnamefont {Gorza}}, \bibinfo {author} {\bibfnamefont {D.}~\bibnamefont
  {Bloch}}, \ and\ \bibinfo {author} {\bibfnamefont {M.}~\bibnamefont
  {Ducloy}},\ }\href {\doibase 10.1209/0295-5075/77/54001} {\bibfield
  {journal} {\bibinfo  {journal} {Europhysics Letters (EPL)}\ }\textbf
  {\bibinfo {volume} {77}},\ \bibinfo {pages} {54001} (\bibinfo {year}
  {2007})}\BibitemShut {NoStop}%
\bibitem [{\citenamefont {Schwarzkopf}\ \emph {et~al.}(2011)\citenamefont
  {Schwarzkopf}, \citenamefont {Sapiro},\ and\ \citenamefont
  {Raithel}}]{schwarzkopf_imaging_2011}%
  \BibitemOpen
  \bibfield  {author} {\bibinfo {author} {\bibfnamefont {A.}~\bibnamefont
  {Schwarzkopf}}, \bibinfo {author} {\bibfnamefont {R.~E.}\ \bibnamefont
  {Sapiro}}, \ and\ \bibinfo {author} {\bibfnamefont {G.}~\bibnamefont
  {Raithel}},\ }\href {\doibase 10.1103/PhysRevLett.107.103001} {\bibfield
  {journal} {\bibinfo  {journal} {Phys. Rev. Lett.}\ }\textbf {\bibinfo
  {volume} {107}},\ \bibinfo {pages} {103001} (\bibinfo {year}
  {2011})}\BibitemShut {NoStop}%
\bibitem [{\citenamefont {Schauß}\ \emph {et~al.}(2012)\citenamefont
  {Schauß}, \citenamefont {Cheneau}, \citenamefont {Endres}, \citenamefont
  {Fukuhara}, \citenamefont {Hild}, \citenamefont {Omran}, \citenamefont
  {Pohl}, \citenamefont {Gross}, \citenamefont {Kuhr},\ and\ \citenamefont
  {Bloch}}]{schaus_observation_2012}%
  \BibitemOpen
  \bibfield  {author} {\bibinfo {author} {\bibfnamefont {P.}~\bibnamefont
  {Schauß}}, \bibinfo {author} {\bibfnamefont {M.}~\bibnamefont {Cheneau}},
  \bibinfo {author} {\bibfnamefont {M.}~\bibnamefont {Endres}}, \bibinfo
  {author} {\bibfnamefont {T.}~\bibnamefont {Fukuhara}}, \bibinfo {author}
  {\bibfnamefont {S.}~\bibnamefont {Hild}}, \bibinfo {author} {\bibfnamefont
  {A.}~\bibnamefont {Omran}}, \bibinfo {author} {\bibfnamefont
  {T.}~\bibnamefont {Pohl}}, \bibinfo {author} {\bibfnamefont {C.}~\bibnamefont
  {Gross}}, \bibinfo {author} {\bibfnamefont {S.}~\bibnamefont {Kuhr}}, \ and\
  \bibinfo {author} {\bibfnamefont {I.}~\bibnamefont {Bloch}},\ }\href
  {\doibase 10.1038/nature11596} {\bibfield  {journal} {\bibinfo  {journal}
  {Nature}\ }\textbf {\bibinfo {volume} {491}},\ \bibinfo {pages} {87}
  (\bibinfo {year} {2012})}\BibitemShut {NoStop}%
\bibitem [{\citenamefont {McQuillen}\ \emph {et~al.}(2013)\citenamefont
  {McQuillen}, \citenamefont {Zhang}, \citenamefont {Strickler}, \citenamefont
  {Dunning},\ and\ \citenamefont {Killian}}]{mcquillen_imaging_2013}%
  \BibitemOpen
  \bibfield  {author} {\bibinfo {author} {\bibfnamefont {P.}~\bibnamefont
  {McQuillen}}, \bibinfo {author} {\bibfnamefont {X.}~\bibnamefont {Zhang}},
  \bibinfo {author} {\bibfnamefont {T.}~\bibnamefont {Strickler}}, \bibinfo
  {author} {\bibfnamefont {F.~B.}\ \bibnamefont {Dunning}}, \ and\ \bibinfo
  {author} {\bibfnamefont {T.~C.}\ \bibnamefont {Killian}},\ }\href {\doibase
  10.1103/PhysRevA.87.013407} {\bibfield  {journal} {\bibinfo  {journal} {Phys.
  Rev. A}\ }\textbf {\bibinfo {volume} {87}},\ \bibinfo {pages} {013407}
  (\bibinfo {year} {2013})}\BibitemShut {NoStop}%
\bibitem [{\citenamefont {Schwarzkopf}\ \emph {et~al.}(2013)\citenamefont
  {Schwarzkopf}, \citenamefont {Anderson}, \citenamefont {Thaicharoen},\ and\
  \citenamefont {Raithel}}]{schwarzkopf_spatial_2013}%
  \BibitemOpen
  \bibfield  {author} {\bibinfo {author} {\bibfnamefont {A.}~\bibnamefont
  {Schwarzkopf}}, \bibinfo {author} {\bibfnamefont {D.~A.}\ \bibnamefont
  {Anderson}}, \bibinfo {author} {\bibfnamefont {N.}~\bibnamefont
  {Thaicharoen}}, \ and\ \bibinfo {author} {\bibfnamefont {G.}~\bibnamefont
  {Raithel}},\ }\href {\doibase 10.1103/PhysRevA.88.061406} {\bibfield
  {journal} {\bibinfo  {journal} {Phys. Rev. A}\ }\textbf {\bibinfo {volume}
  {88}},\ \bibinfo {pages} {061406} (\bibinfo {year} {2013})}\BibitemShut
  {NoStop}%
\bibitem [{\citenamefont {Robicheaux}(2005)}]{robicheaux_ionization_2005}%
  \BibitemOpen
  \bibfield  {author} {\bibinfo {author} {\bibfnamefont {F.}~\bibnamefont
  {Robicheaux}},\ }\href {\doibase 10.1088/0953-4075/38/2/024} {\bibfield
  {journal} {\bibinfo  {journal} {J. Phys. B: At. Mol. Opt. Phys.}\ }\textbf
  {\bibinfo {volume} {38}},\ \bibinfo {pages} {S333} (\bibinfo {year}
  {2005})}\BibitemShut {NoStop}%
\bibitem [{\citenamefont {Reinhard}\ \emph {et~al.}(2007)\citenamefont
  {Reinhard}, \citenamefont {Liebisch}, \citenamefont {Knuffman},\ and\
  \citenamefont {Raithel}}]{reinhard_level_2007}%
  \BibitemOpen
  \bibfield  {author} {\bibinfo {author} {\bibfnamefont {A.}~\bibnamefont
  {Reinhard}}, \bibinfo {author} {\bibfnamefont {T.~C.}\ \bibnamefont
  {Liebisch}}, \bibinfo {author} {\bibfnamefont {B.}~\bibnamefont {Knuffman}},
  \ and\ \bibinfo {author} {\bibfnamefont {G.}~\bibnamefont {Raithel}},\ }\href
  {\doibase 10.1103/PhysRevA.75.032712} {\bibfield  {journal} {\bibinfo
  {journal} {Phys. Rev. A}\ }\textbf {\bibinfo {volume} {75}},\ \bibinfo
  {pages} {032712} (\bibinfo {year} {2007})}\BibitemShut {NoStop}%
\bibitem{supplement} See Supplemental Material for calculation results.
\bibitem [{\citenamefont {Möbius}\ \emph {et~al.}(2013)\citenamefont
  {Möbius}, \citenamefont {Genkin}, \citenamefont {Wüster}, \citenamefont
  {Eisfeld},\ and\ \citenamefont {Rost}}]{mobius_breakup_2013}%
  \BibitemOpen
  \bibfield  {author} {\bibinfo {author} {\bibfnamefont {S.}~\bibnamefont
  {Möbius}}, \bibinfo {author} {\bibfnamefont {M.}~\bibnamefont {Genkin}},
  \bibinfo {author} {\bibfnamefont {S.}~\bibnamefont {Wüster}}, \bibinfo
  {author} {\bibfnamefont {A.}~\bibnamefont {Eisfeld}}, \ and\ \bibinfo
  {author} {\bibfnamefont {J.~M.}\ \bibnamefont {Rost}},\ }\href {\doibase
  10.1103/PhysRevA.88.012716} {\bibfield  {journal} {\bibinfo  {journal} {Phys.
  Rev. A}\ }\textbf {\bibinfo {volume} {88}},\ \bibinfo {pages} {012716}
  (\bibinfo {year} {2013})}\BibitemShut {NoStop}%
\end{thebibliography}

%

\newpage
\clearpage
\section*{Supplemental Material}
We have used the model presented in detail in [$i$] to calculate the interaction potential between a pair of $70S_{1/2}$ rubidium Rydberg atoms with dipole-dipole interaction terms only, as well as with dipole-dipole, dipole-quadrupole and quadrupole-quadrupole terms. The projection of the electron angular momenta onto the internuclear axis, M, can take the values 0 or $\pm1$. For M=0 there are two Rydberg-pair $70S_{1/2}-70S_{1/2}$  levels, labeled A and B, while for M=$\pm1$ there is only one (see table). QQ means all interactions up to and including quadrupole-quadrupole terms are included. DD means only dipole-dipole terms are included. 

It is seen that the atom pairs have very similar shifts on all potentials, and that interaction terms beyond the dipole-dipole term are not important in the atom distance range of interest (R$>$7~$\mu$m). The results also agree well with a perturbative level-shift calculation in [$ii$].

\begin{table}[h]
    \centering{
	\setlength{\tabcolsep}{4pt}
    \begin{tabular}{| c | c | c | c | c | c | c |} 
    \hline 		
 &  \multicolumn{2}{|c|}{M=0 Level A} & \multicolumn{2}{c|}{M=0 Level B} & \multicolumn{2}{c|}{M=1}\\	
\hline
R	&	QQ	&	DD	&	QQ	&	DD	&	QQ	&	DD	\\
$\mu$m	&	MHz	&	MHz	&	MHz	&	MHz	&	MHz	&	MHz	\\
\hline
6	&	18.215	&	18.223	&	19.046	&	19.052	&	18.424	&	18.431	\\
7	&	7.263	&	7.268	&	7.604	&	7.609	&	7.349	&	7.353	\\
8	&	3.267	&	3.269	&	3.421	&	3.424	&	3.305	&	3.307	\\
9	&	1.613	&	1.614	&	1.69	&	1.691	&	1.632	&	1.633	\\
10	&	0.858	&	0.858	&	0.898	&	0.899	&	0.868	&	0.868	\\
\hline

    \hline 
    \end{tabular}}
    \label{table:potential}
\end{table}

\noindent\\
$\left[i\right]$ J. Deiglmayr, H. Saßmannshausen, P. Pillet, and F. Merkt, Phys. Rev. Lett. 113, 193001 (2014).\\
$\left[ii\right]$ A. Reinhard, T. C. Liebisch, B. Knuffman, and G. Raithel, Phys. Rev. A 75, 032712 (2007).\\

\end{document}